\pdfoutput=5
\documentclass[11pt]{article}

\usepackage[review]{acl}

\usepackage{times}
\usepackage{latexsym}
\usepackage{amsmath}
\usepackage{enumitem}
\usepackage{graphicx}
\usepackage{array} 
\usepackage{booktabs}

\usepackage[T1]{fontenc}
\usepackage[utf8]{inputenc}

\usepackage{microtype}

\newcommand{\mc}[1]{\mathcal{#1}}
\hypersetup{nolinks=true}
\title{Using Artificial Intuition in Distinct, Minimalist Classification of Scientific Abstracts for Management of Technology Portfolios\thanks{  
This research was funded in part by the National Science Foundation (NSF) award 2048703.}}

\author{Prateek Ranka and
         Fred Morstatter\\
         Information Sciences Institute, Viterbi School of Engineering, University of Southern California \\ 
         {\bf Alexandra Graddy-Reed} \\ 
         Sol Price School of Public Policy, University of Southern California \\
         {\bf Andrea Belz} \\ 
         Information Sciences Institute, Viterbi School of Engineering, University of Southern California
         }
\begin{document}
\nolinenumbers
%\maketitle
{\makeatletter\acl@finalcopytrue
  \maketitle
}
\begin{abstract}
Classification of scientific abstracts is useful for strategic activities but challenging to automate because the sparse text provides few contextual clues. Metadata associated with the scientific publication can be used to improve performance but still often requires a semi-supervised setting.  Moreover, such schemes may generate labels that lack distinction- namely, they overlap and thus do not uniquely define the abstract.  In contrast, experts label and sort these texts with ease.  Here we describe an application of a process we call ``artificial intuition'' to replicate the expert's approach, using a Large Language Model (LLM) to generate metadata. We use publicly available abstracts from the United States National Science Foundation to create a set of labels, and then we test this on a set of abstracts from the Chinese National Natural Science Foundation to examine funding trends.  We demonstrate the feasibility of this method for research portfolio management, technology scouting, and other strategic activities. 
\end{abstract}

\section{Introduction.}
Management of research and development activities requires understanding the landscape of relevant scientific activities, such as that of a typical literature search where one sorts through scientific  abstracts and evaluates their applicability to the research question. It is infrequently recognized that this thought process is identical to determine funding allocations in a research portfolio or identifying technology scouting candidates.

In each task an abstract is processed, labeled, and sorted, but different strategies follow from the task objective.  Specifically, if one seeks to conduct broad exploration, then an abstract should be \textit{maximally labeled}. For example, a paper on ``biometrics'' would be labeled with ``biology'' and ``data sciences'' so that it can be retrieved regardless of the path to find it.  This approach drives the typical process by which an author generates keywords for a paper. That is, an external taxonomy is proposed and the author assigns \textit{all} the potential classifications (as represented by keywords) to the text.  A longer list of classes increases the chances of discovery. 

On the other hand, simple strategic funding allocations call for \textit{minimal labeling}.  For instance, a biometrics award or grant abstract risks being ``double-counted'' if categorized as both biology and data sciences.  In principle, one can imagine schemes in which the funding is weighted by the number of labels - a biometrics award could be counted half as biology, and half as data sciences.  While this scheme could be correct, it would be more parsimonious to create a simple biometrics label and assign the abstract entirely to this new field.  Importantly, this mimics typical funding processes in organizations like federal agencies: one program manager oversees 100\% of an award. 

Thus, in contrast to broad exploration wherein many search paths - and thus many labels - for a document are desirable, strategic research portfolio management calls for creating a minimalist set of labels so that each abstract is ideally found only once.  To do so, the labels must be \textit{distinct}, with little or no overlap.  

An expert conducts the process of labeling abstracts with ease while accessessing troves of information unavailable to the less informed reader:  the expert reads the abstract and extracts keywords, then retrieves metadata from a library obtained through extensive training. With the benefit of these metadata, the expert can label the abstract either maximally  - namely, find a large number of conceptual paths to the abstract, such as that of a literature search; or minimally - discovering the parsimonious term to encompasses the abstract's nature.  This process is difficult to replicate in an unsupervised system. 

Here we implement \textit{artificial intuition}, a technique in which we construct metadata with a Large Language Model (LLM) to augment the original source data - in effect, replicating the process of an expert \citep{Sakhrani2024}.  Unlike LLM use in  tasks such as matching manuscripts and reviewers \citep{zhang2024paperreviewermatching} or scientific literature processing \citep{zhang2023contrastivelearning-maple}, here we use the LLM to augment the text of a short document.  

As our laboratory, we use the publicly available award abstracts of the US National Science Foundation (US NSF) to build and characterize a model.  We test it by labeling a set of abstracts from the Chinese National Natural Science Foundation and compare our label assignment to manual labels.  

This work has several important implications. In the realm of text analysis, we report on the parsimonious use of LLMs and validate new metrics to characterize models.  We contribute to the field of research management by developing a tool to describe research portfolio allocation at scale.  

\section{Related work.}

We summarize related work that addresses core challenges in scientific text classification, such as data scarcity, domain-specific terminology, and complex taxonomies. Recent advances leverage Large Language Models (LLMs), retrieval-augmented generation (RAG), clustering, and semantic representations to improve classification performance.

% \subsection{LLM-based Data Augmentation Techniques}

LLMs are increasingly used to improve text classification by either paraphrasing existing samples or generating new ones. While many studies use random sample selection for few-shot prompting, recent work has explored more deliberate strategies. \citet{Cegin2024} find that random selection often performs best for in-distribution data, while informed strategies offer only limited and inconsistent benefits, especially for out-of-distribution cases. \citet{Glazkova2024} show that prompt-based augmentation strategies improve performance on multi-label ecological text classification. \citet{Zhao2024} compare rewriting and new sample generation using ChatGPT and find that the latter is generally more effective. 

% \subsection{RAG-based Methods}
% AB TO START REVIEWING HERE

To address limited labeled data in scientific text classification, \citet{Zhong2025Clustering} propose a technique, which clusters data to select representative samples for labeling and uses them for RAG, LLM rewriting, and synonym substitution.
% This reduces labeling effort and boosts accuracy. 
\citet{Jeong2024AutomaticClassification} present a RAG-enhanced LLM system that retrieves document context from a vector database to improve classification, outperforming BERT models and even manual labels on small datasets. \citet{Li2024KRA} introduce a k-NN-based RAG method that augments top retrieved neighbors and incorporates them into predictions. 
% It improves accuracy for BERT and RoBERTa, especially on short texts, with ablation studies confirming the value of combining retrieval and augmentation.

Hierarchical Text Classification (HTC) is challenging due to complex taxonomies, limited labeled data, and sparse context in short texts. \citet{Longola2024HTC-GEN}  uses LLMs to generate synthetic data across hierarchy levels, improving performance and reducing annotation needs. \citet{Zhang2025TELEClass} introduce a minimally supervised method that enriches taxonomies with class-indicative terms and uses LLMs for efficient annotation and document generation, improving coverage of fine-grained classes. 

% \citep{Stein2019WordEmbeddingsHTC} show that word embeddings and hierarchical strategies (e.g., fastText with LCPN) outperform flat approaches, underscoring the value of embeddings and structure-aware models for HTC.

% \subsection{LLM-based Ranking Techniques}
Scientific paper retrieval is challenged by domain-specific language, limited labels, and dense retrievers missing fine-grained concepts.  \citet{Zhang2025SemRank} introduce a concept-based semantic indexing and reranking method guided by LLMs. \citet{Tian2025CoRank} present a training-free, model-agnostic reranker that first uses compact features for broad coverage, then applies full-text reranking. Both approaches improve fine-grained retrieval while overcoming token and data limitations.

% \subsection{Other Approaches}

\citet{Sinoara2019} generate semantically rich, low-dimensional document vectors using word sense disambiguation and embedded representations. 
% These methods outperform traditional BOW and LDA models in classification tasks, offering better interpretability and efficiency, especially for semantically complex texts. 
\citet{Ni2020} propose a keyword-based classification method using automatic keyword extraction and genetic algorithm optimization. Their models show improved accuracy and F-Measure on Association for Computing Machinery (ACM) datasets, supported by a self-updating keyword model for continuous learning.

% \FIX{Write a summary statement that explains why our method is necessary.}

While recent advances leverage LLMs, RAG, and clustering to improve scientific text classification accuracy, these approaches primarily focus on fitting texts into existing taxonomies or maximizing retrieval paths. Our approach differs fundamentally by creating minimalist, orthogonal label spaces that mirror how experts categorize abstracts for strategic decisions. By optimizing LLM-augmented keyword extraction, we generate parsimonious label sets that avoid the double-counting inherent in multi-label schemes. This addresses a critical gap in research portfolio management, where unique categorization is essential for accurate funding allocation and strategic analysis.

% \textcolor{red}{TODO: Wrap up paragraph.}

\section{Approach.}

% Write a paragraph summarizing the approach from Naseela's paper

\begin{figure*}
    \centering
    \includegraphics[width=1.0\linewidth]{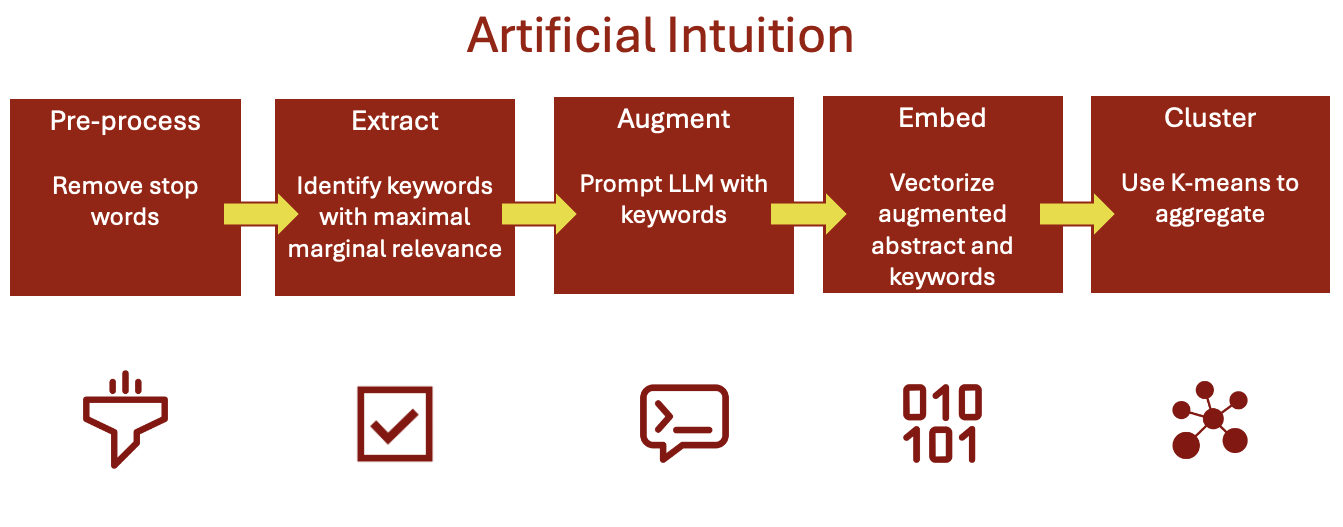}
    \caption{Artificial intuition process.}
    \label{fig:art.int}
\end{figure*}

The method proposed by \citet{Sakhrani2024} automates the task of classifying short scientific texts, a task that is inherently difficult due to the limited context and semantic cues in such documents. Their approach introduces the concept of \textit{artificial intuition}, which seeks to replicate the rapid, knowledge-informed classification typically performed by human experts, rather than relying on existing, often inconsistent, taxonomies.  The artificial intuition method proceeds as follows (Figure \ref{fig:art.int}): 

\begin{itemize}
    \item Extract keywords from corpus abstracts using YAKE.
    \item Augment extracted keywords with contextual data using the Gemini-2.0-Flash LLM.
    \item Apply Maximal Marginal Relevance (MMR) to refine the keyword embeddings, balancing relevance and novelty.
    \item Cluster the resulting augmented keyword embeddings using K-Means.
\end{itemize}

\section{Analysis.}

\subsection{Training and test corpora.}
As a training corpus, we downloaded 3,865 publicly available abstracts for US NSF awards in California made from September 1, 2017 to October 1, 2022. As we sought to characterize abstracts in the basic sciences and engineering, we excluded awards in the STEM Education (EDU) and Social, Behavioral, and Economic Sciences (SBE) Directorates and in the Translational Impacts (TI) Division. 3,033 abstracts remained for consideration.  With 200 reserved for internal tests, 2,833 were used for training.

We created a comprehensive pre-processing pipeline to prepare abstracts for keyword extraction. Proper nouns such as names, universities, and organizations are removed using regex patterns, POS tagging, and NER, as they often introduce noise and skew keyword prioritization. To ensure representativeness, we filter out terms that are rare (appearing in $< 10\%$ of the corpus) or frequent (in $>75\%$ of documents). A curated stopword list removes abbreviations, articles, prepositions, and other non-semantic tokens. We remove URLs, special characters, non-ASCII text, and number words, and conduct case normalization and artifact filtering. This results in semantically rich, clean text for unbiased keyword extraction.

Before pre-processing the average abstract length was 414 words; after, it was 166 words with a minimum and maximum length of 87 and 405 words, respectively. We limited keywords as follows: for abstracts with $<60$ tokens, we extracted 1 keyword; at 60-85 tokens, 2 keywords; and 3 keywords for abstracts of length $>85$ tokens.  Consequently, 8,499 unique keyword phrases, typically trigrams, were considered.

As a test data set, we downloaded 367 abstracts of work funded by the National Natural Science Foundation of China (NSFC) analyzed by \citet{Liu2019NSFCAnalysis}, of average length 172 words.  After pre-processing, the test data set had an average length of 81 words with a minimum (maximum) of 22 (168) words.

\subsection{Extracted keywords.}
\begin{table*}[ht!]
\centering
\scriptsize
\caption{Keywords for $\hat{k}$ = 6 with four seeds. }
\vspace{0.5em}
\begin{tabular}{cccc}
\toprule
\textbf{Seed 1} & \textbf{Seed 2} & \textbf{Seed 3} & \textbf{Seed 4} \\
\midrule
% industries government labs & industries government labs & industries government labs & industries government labs \\
global environmental change & global environmental change & -- & -- \\
-- & -- & ecology & ecology \\
cellular behavior long-term & -- & -- & -- \\
-- & efforts understand disease & -- & -- \\
-- & -- & biology materials technology & biology materials technology \\
theoretical physics include & theoretical physics include & theoretical physics include & theoretical physics include \\
materials nanostructures theoretical & materials nanostructures theoretical & materials nanostructures theoretical & materials nanostructures theoretical \\
networked systems crucial & networked systems crucial & -- & -- \\
-- & -- & disciplines participants nsf-funded & disciplines participants nsf-funded \\
\bottomrule
\end{tabular}
\label{tab:seed_keywords.6}
\end{table*}

\begin{table*}[ht!]
\centering
\scriptsize
\caption{Keywords for $\hat{k}$ = 10 with four seeds.}
\vspace{0.5em}
\begin{tabular}{cccc}
\toprule
\textbf{Seed 1} & \textbf{Seed 2} & \textbf{Seed 3} & \textbf{Seed 4} \\
\midrule
networked systems crucial            & networked systems crucial            & networked systems crucial            & networked systems crucial \\
--      & --                                   & natural promising applications       & natural promising applications \\

crustal environment fluid-rock       & crustal environment fluid-rock       & --      & -- \\
--                                   & --                                   & --                                   & estimating annual subduction \\
ecology                              & ecology                              & ecology                              & ecology \\
sea ice loss & -- & -- & --\\
--                                   & flooding wildfires hurricanes        & --                                   & -- \\

holes uncovering astrophysical       & --                                   & holes uncovering astrophysical       & -- \\
% industries government labs           & industries government labs           & industries government labs           & industries government labs \\
mathematical sciences including      & --                                   & mathematical sciences including      & -- \\
print polymers mechanical            & print polymers mechanical            & print polymers mechanical            & print polymers mechanical \\
molecular-level interactions researchers & --                               & --                                   & -- \\
--                                   & --                                   & cellular behavior long-term          & cellular behavior long-term \\
materials nanostructures theoretical & materials nanostructures theoretical & --                                   & materials nanostructures theoretical \\
--                                   & macromolecular processes motivated   & --                                   & -- \\
--                                   & --                                   & nanoscale electronic structure       & -- \\
--                                   & theoretical physics include          & --                                   & theoretical physics include \\
--                                   & --                                   & disciplines participants nsf-funded  & -- \\
--                                   & --                                   & --                                   & engaging scientists collaborating \\
--                                   & scientists university-based researchers & --                                & -- \\
\bottomrule
\end{tabular}
\label{tab:seed_keywords.10}
\end{table*}

We ran the artificial intuition method to generate a label set for the 2,833 abstracts described above.  Each cluster was labeled with the keyword closest to its centroid.  As examples, we report the extracted cluster labels for $\hat{k}$ = 6 and $\hat{k}$ = 10 for four distinct seeds, to show the types of keywords and the stability of the system when the initial parameterization varies (Tables \ref{tab:seed_keywords.6} and \ref{tab:seed_keywords.10}). In both cases, the system created a label of ``industries government labs'', but we exclude this label as a generic artifact of the NSF abstract corpus.  Therefore, the reported label sets have one fewer keyword than the input parameter of $\hat{k}$.  This allows us to focus on engineering and scientific document classification.  In general, the words appear to cover several fields of broad scientific interest.  

\subsection{Model characterization.}
\citet{Sakhrani2024} proposed two metrics to characterize the quality of the labeling scheme:  Redundancy $\mc{R}$ to describe the independence of the labels, and coverage $\mc{S}$ to describe how completely the label set describes the corpus. 

We define redundancy $\mc{R}$ as the maximum cosine similarity between any two label vectors in the label space and is thus bounded by 0 and 1; small values indicate that the labels are nearly orthogonal, whereas values closer to 1 indicates that the labels overlap.  Formally, for normalized label vectors $\mathbf{T}_i$ and $\mathbf{T}_j$ in a label space $\mc{L}$, redundancy $\mc{R}$ is the maximum cosine similarity for all pairs of labels:
\begin{equation}
\mathcal{R} = \max_{i \neq j} \left( \mathrm{cosine\ similarity}(\mathbf{T}_i, \mathbf{T}_j) \right)
\end{equation}

\begin{equation}
\mathrm{cosine\ similarity}(\mathbf{T}_i, \mathbf{T}_j) = \frac{\mathbf{T}_i \cdot \mathbf{T}_j}{\|\mathbf{T}_i\| \|\mathbf{T}_j\|}
\end{equation}

The coverage $\mc{S}$ for a document corpus $\mathcal{D}$ is less intuitive but can also be conceived geometrically.  A document's $\hat{c}$ keywords are embedded with dimensionality $v$ (768 in our case) such that the document is described by a matrix $\mathcal{C}$ of dimensions $v \times \hat{c}$.  We create a label embedding matrix $\mathcal{L}$ with dimensions $\hat{k} \times v$ for $\hat{k}$ generated labels.

To quantify the semantic alignment between the label space and the enriched keyword space, we define the \textit{coverage matrix} $\mathcal{W}$, with elements computed as:
\begin{equation}
w_{ij} = \sum_v \mathcal{L}_{iv} \mathcal{C}_{vj}
\end{equation}
Here, $w_{ij}$ quantifies the alignment between label $i$ and keyword $j$, effectively projecting the document’s semantic components onto the label space. Each $w_{ij}$ ranges between -1 and 1, reflecting the cosine similarity of embeddings in a normalized high-dimensional space.

The coverage $S^d$ for a specific document $d$ is determined by identifying the maximal alignment across its associated keywords:
\begin{equation}
S^d = \max(w_{ij}^d)
\end{equation}
This formulation captures the best match between the document’s semantic content and the available labels, highlighting the extent to which the document is represented within the label space. $\mc{S}$ increases with $\hat{k}$ and levels off when the the labels represent the documents relatively well. 

To compute the overall corpus-level coverage $S^D$, we average the document-level coverages across all documents in the corpus $\mathcal{D}$:
\begin{equation}
S^D = \frac{\sum_D{\mathcal{S}}}{\mathcal{D}}
\end{equation}

Our model is thus optimized if the redundancy $\mc{R}$ is low but the coverage $\mc{S}$ is high. 

\begin{figure}[h!]
    \centering
    \includegraphics[width=1.0\linewidth]{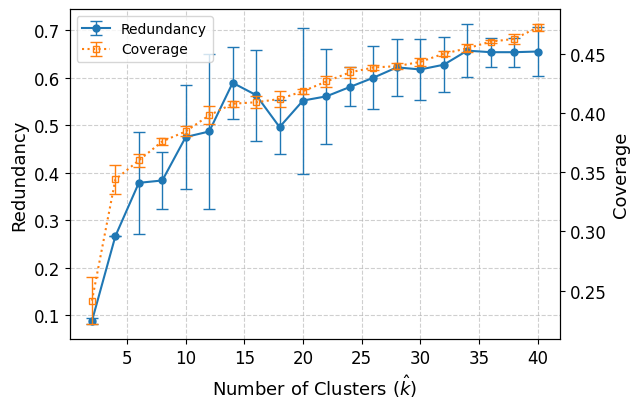}
    \caption{Variation of redundancy $\mc{R}$ (blue solid line) and coverage $\mc{S}$ with clusters $\hat{k}$ (orange dotted line).  Averages and errors are calculated across three distinct seeds.}
    \label{fig:cov_and_red}
\end{figure}

We characterized the redundancy $\mathcal{R}$ and coverage $\mathcal{S}$ in the range $\hat{k}$ = 2-40 by sampling the data set with four different seeds (Figure \ref{fig:cov_and_red}).  $\mc{R}$ shows reaches an asymptotic value at about $\hat{k}=15$ and more clusters do not add to the information.  On the other hand, $\mc{S}$ increases consistently and appears to be less noisy.  Higher values of $\hat{k}$ significantly increase the coverage, with the $\hat{k} = 6$ serving as a minimum for performance. We therefore opt to test our model at this value to keep the $\mc{R}$ low. 

\subsection{Labeling the test data set. }
\begin{figure}
    \centering
    \includegraphics[width=1.0\linewidth]{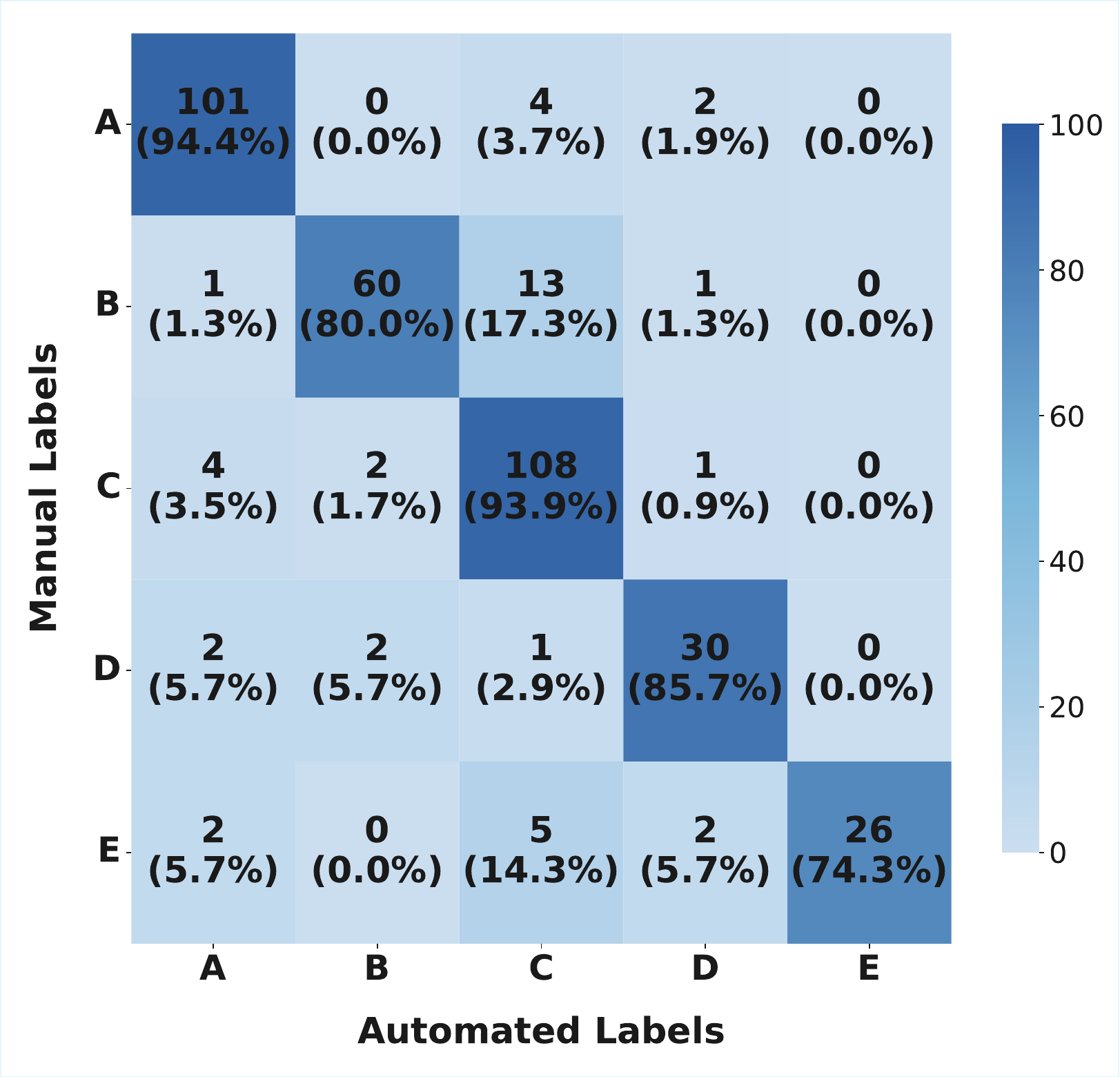}
    \caption{Confusion matrix comparing the manual and automated labeling methods.   Percentages are normalized by row to show the agreement of the automated method.  Labels:  A = cellular behavior long-term, B = global environmental change, C = materials nanostructures theoretical, D = networked systems crucial, E = theoretical physics include.}
    \label{fig:label_confusion_matrix}
\end{figure}

We turn our attention to the test data set of abstracts from the NSFC, labeled with our artificial intuition system and manually for comparison.  We report performance in a confusion matrix (Figure \ref{fig:label_confusion_matrix}) and demonstrate strong performance overall, with over 75\% accuracy in four of the five labeled topics. One category shows notable confusion patterns. Abstracts manually labeled as ``theoretical physics include'' are frequently misclassified as ``materials nanostructures theoretical'' (14.3\% misclassification), suggesting conceptual overlap between these fields where materials science and theoretical physics converge.

\begin{figure*}
    \centering
    \includegraphics[width=1.0\linewidth]{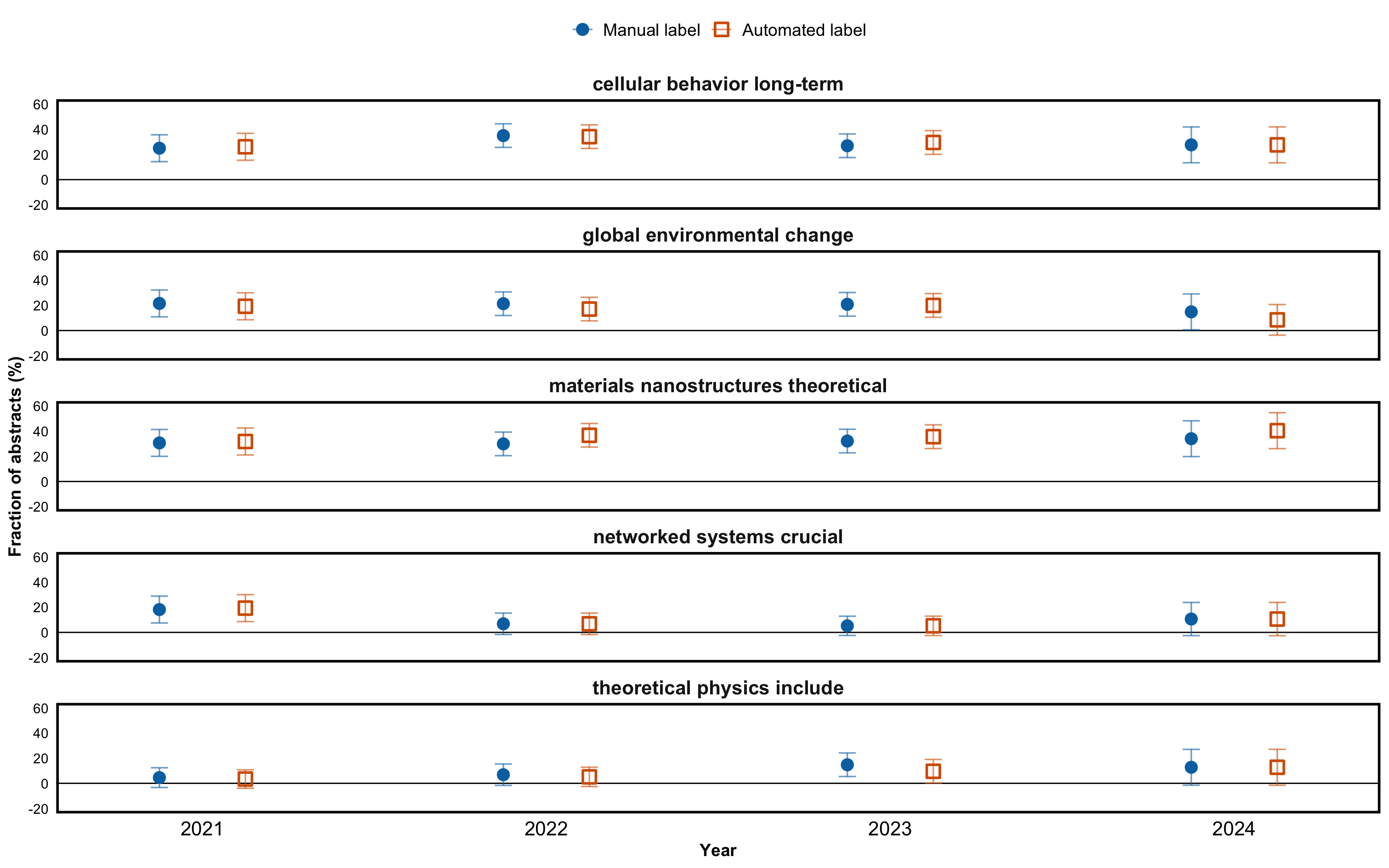}
    \caption{Evolution of the portfolio of abstracts reported by the National Natural Science Foundation of China, measured with both manual (blue circle) and automated (orange square) methods.}
    \label{fig:labels_error}
\end{figure*}

We then demonstrate an application of artificial intuition by analyzing the test data set with both manual and automated methods (Figure \ref{fig:labels_error}), where the reported error bars describe 95\% multinomial confidence intervals.  Consistent with Figure \ref{fig:label_confusion_matrix}, we show general agreement between the manual and automated methods, although the automated process estimates a slightly higher concentration in the biological sciences (``cellular behavior long-term'') that is compensated in the data sciences (``networked systems crucial'').  

More importantly, we observe funding trends.  For instance, we report here that the activity (as measured by the number of abstracts) in materials and nanostructures represented about 40\% of the total, whereas theoretical physics represented only a few percent.  Likewise, we document a high level of activity in ``materials nanostructures theoretical''.  

\section{Discussion and Future Research.}
To our knowledge, this is one of the first studies to  automate abstract labeling for research strategy.  In the scientific document processing community, less attention has been paid to minimalist schemes for strategic analysis. We compensate for each document's brevity by augmenting it with additional text generated by an LLM.  This leads to several important findings. 

First, we report that this method generates labels that roughly track with the portfolio of a broad scientific enterprise like the NSF.  The previous work by Sakhrani et al. analyzed a set of aerospace abstracts.  Here we demonstrate a higher level of classification across a broader set of topics.  One can imagine running this iteratively to create tiers of precision in the labeling process. 

Second, we provide new insight into the measures previously proposed by Sakhrani et al. Our tests demonstrate that the redundancy is noisier than the coverage, but both behave as expected and can be used as guides to optimize the selection of the number of clusters.  This could be further automated by fitting a simple exponential growth model to the data and selecting an optimum (for instance, analyzing the value of $\hat{k}$ where each curve reaches 75\% of their asymptotic value).  

Third, we show strong agreement between the automatic classification and a manual process, as evidenced by the confusion matrix.  Further work is needed to understand the the off-diagonal elements.  

Finally, we provide new evidence that artificial intuition can be used for strategic management.  This opens the door to many new avenues of research.  For instance, one could follow the authors and create an entirely new network model that describes the evolution of scientific research.  Alternatively, it could be possible to track and even predict how science is transformed into invention by identifying emerging fields.  
 
\section*{Acknowledgments.}
We thank Rithik Kompelli and Akash Mittapally for contributions and Charles Parker for support.

\bibliography{anthology, custom, artificial.intuition}

\begin{thebibliography}{16}
\expandafter\ifx\csname natexlab\endcsname\relax\def\natexlab#1{#1}\fi

\bibitem[{Cegin et~al.(2024)Cegin, Pecher, Simko, Srba, Bieliková, and Brusilovsky}]{Cegin2024}
Ján Cegin, Branislav Pecher, Jakub Simko, Ivan Srba, Mária Bieliková, and Peter Brusilovsky. 2024.
\newblock \href {https://doi.org/10.48550/arXiv.2410.10756} {Use random selection for now: Investigation of few-shot selection strategies in llm-based text augmentation for classification}.
\newblock \emph{CoRR}, abs/2410.10756.

\bibitem[{Glazkova and Zakharova(2024)}]{Glazkova2024}
Anna Glazkova and Olga Zakharova. 2024.
\newblock \href {https://doi.org/10.1109/ISPRAS64596.2024.10899128} {Evaluating llm prompts for data augmentation in multi-label classification of ecological texts}.
\newblock In \emph{2024 Ivannikov Ispras Open Conference (ISPRAS)}, pages 1--7.

\bibitem[{Jeong and Jin(2024)}]{Jeong2024AutomaticClassification}
Jaehan Jeong and Dongsup Jin. 2024.
\newblock \href {https://doi.org/10.56977/jicce.2024.22.4.280} {Automatic classification of scientific and technical papers using large language models and retrieval-augmented generation}.
\newblock \emph{J. Inf. Commun. Converg. Eng.}, 22(4):280--287.

\bibitem[{Li et~al.(2024)Li, Tang, Lei, Zhang, Li, Yu, Pi, and Hu}]{Li2024KRA}
Jie Li, Chang Tang, Zhechao Lei, Yirui Zhang, Xuan Li, Yanhua Yu, Renjie Pi, and Linmei Hu. 2024.
\newblock \href {https://doi.org/10.3390/electronics13163237} {Kra: K-nearest neighbor retrieval augmented model for text classification}.
\newblock \emph{Electronics}, 13(16):3237.

\bibitem[{Liu et~al.(2019)Liu, Gao, Wang, Wang, Shen, and Wang}]{Liu2019NSFCAnalysis}
Y.~Liu, Z.~Gao, H.~Wang, J.~Wang, J.~Shen, and C.~Wang. 2019.
\newblock \href {https://doi.org/10.21037/atm.2019.05.63} {Analysis of projects funded by the national natural science foundation of china during the years of 2014–2018}.
\newblock \emph{Annals of Translational Medicine}, 7(12):267.

\bibitem[{Longola et~al.(2024)Longola, Mongiovì, Bulla, and Tuccari}]{Longola2024HTC-GEN}
Carmelo~Fabio Longola, Misael Mongiovì, Luana Bulla, and Giusy~Giulia Tuccari. 2024.
\newblock \href {https://doi.org/10.5220/0012790700003756} {Htc-gen: A generative llm-based approach to handle data scarcity in hierarchical text classification}.
\newblock In \emph{Proceedings of the 13th International Conference on Data Science, Technology and Applications (DATA 2024)}, pages 129--138. SCITEPRESS-Science and Technology Publications, Lda.

\bibitem[{Ni et~al.(2020)Ni, Li, and Chang}]{Ni2020}
Pin Ni, Yuming Li, and Victor Chang. 2020.
\newblock \href {https://doi.org/10.4018/IJEIS.2020100101} {Research on text classification based on automatically extracted keywords}.
\newblock \emph{International Journal of Enterprise Information Systems}, 16(4):1–16.

\bibitem[{Sakhrani et~al.(2024)Sakhrani, Pervez, Ravikumar, Morstatter, Graddy-Reed, and Belz}]{Sakhrani2024}
Harsh Sakhrani, Naseela Pervez, Anirudh Ravikumar, Fred Morstatter, Alexandra Graddy-Reed, and Andrea Belz. 2024.
\newblock \href {https://aclanthology.org/2024.sdp-1.18/} {Artificial intuition: Efficient classification of scientific abstracts}.
\newblock In \emph{Proceedings of the Fourth Workshop on Scholarly Document Processing (SDP 2024)}, pages 191--201, Bangkok, Thailand. Association for Computational Linguistics.

\bibitem[{Sinoara et~al.(2019)Sinoara, Camacho-Collados, Rossi, Navigli, and Rezende}]{Sinoara2019}
Roberta~A. Sinoara, Jose Camacho-Collados, Rafael~G. Rossi, Roberto Navigli, and Solange~O. Rezende. 2019.
\newblock \href {https://doi.org/10.1016/j.knosys.2018.10.026} {Knowledge-enhanced document embeddings for text classification}.
\newblock \emph{Knowledge-Based Systems}, 163:955--971.

\bibitem[{Tian et~al.(2025)Tian, Xu, Jin, Kang, and Han}]{Tian2025CoRank}
Runchu Tian, Xueqiang Xu, Bowen Jin, SeongKu Kang, and Jiawei Han. 2025.
\newblock \href {https://arxiv.org/abs/2505.13757} {Llm-based compact reranking with document features for scientific retrieval}.
\newblock \emph{arXiv:2505.13757v1 [cs.IR]}.

\bibitem[{Zhang et~al.(2023)Zhang, Cheng, Shen, Liu, Wang, and Gao}]{zhang2023contrastivelearning-maple}
Yu~Zhang, Hao Cheng, Zhihong Shen, Xiaodong Liu, Ye-Yi Wang, and Jianfeng Gao. 2023.
\newblock \href {https://doi.org/10.18653/v1/2023.findings-emnlp.820} {Pre-training multi-task contrastive learning models for scientific literature understanding}.
\newblock In \emph{Findings of the Association for Computational Linguistics: EMNLP 2023}, pages 12259--12275, Singapore. Association for Computational Linguistics.

\bibitem[{Zhang et~al.(2025{\natexlab{a}})Zhang, Shen, Kang, Chen, Jin, and Han}]{zhang2024paperreviewermatching}
Yu~Zhang, Yanzhen Shen, SeongKu Kang, Xiusi Chen, Bowen Jin, and Jiawei Han. 2025{\natexlab{a}}.
\newblock Chain-of-factors paper-reviewer matching.
\newblock In \emph{WWW'25}.

\bibitem[{Zhang et~al.(2025{\natexlab{b}})Zhang, Yang, Jiao, Kang, and Han}]{Zhang2025SemRank}
Yunyi Zhang, Ruozhen Yang, Siqi Jiao, SeongKu Kang, and Jiawei Han. 2025{\natexlab{b}}.
\newblock \href {https://arxiv.org/abs/2505.21815} {Scientific paper retrieval with llm-guided semantic-based ranking}.
\newblock \emph{arXiv:2505.21815v1 [cs.IR]}.

\bibitem[{Zhang et~al.(2025{\natexlab{c}})Zhang, Yang, Xu, Li, Xiao, Shen, and Han}]{Zhang2025TELEClass}
Yunyi Zhang, Ruozhen Yang, Xueqiang Xu, Rui Li, Jinfeng Xiao, Jiaming Shen, and Jiawei Han. 2025{\natexlab{c}}.
\newblock Teleclass: Taxonomy enrichment and llm-enhanced hierarchical text classification with minimal supervision.
\newblock \emph{WWW '25}, pages 2032--2042.

\bibitem[{Zhao et~al.(2024)Zhao, Chen, Ruggles, Feng, Singh, and Yoon}]{Zhao2024}
Huanhuan Zhao, Haihua Chen, Thomas~A. Ruggles, Yunhe Feng, Debjani Singh, and Hong-Jun Yoon. 2024.
\newblock \href {https://doi.org/10.3390/electronics13132535} {Improving text classification with large language model-based data augmentation}.
\newblock \emph{Electronics}, 13(13):2535.

\bibitem[{Zhong et~al.(2025)Zhong, Zeng, Yu, and Lin}]{Zhong2025Clustering}
Shan Zhong, Jiahao Zeng, Yongxin Yu, and Bohong Lin. 2025.
\newblock \href {https://doi.org/10.1007/s41060-025-00774-3} {Clustering algorithms and rag enhancing semi-supervised text classification with large llms}.
\newblock \emph{International Journal of Data Science and Analytics}.

\end{thebibliography}
\bibliographystyle{acl_natbib}

\end{document}